\documentclass[epj,referee]{svjour}
\usepackage{graphicx}
\begin{document}
\title{Extraction of physical laws from joint experimental data}
\author{Igor Grabec}
\institute{Faculty of Mechanical Engineering, University of Ljubljana,\\
A\v{s}ker\v{c}eva 6, PP 394, 1001 Ljubljana, Slovenia,\\ 
Tel: +386 01 4771 605, Fax: +386 01 4253 135, \\
E-mail: igor.grabec@fs.uni-lj.si \\}
\date{Received: date / Revised version: date}

\abstract{The extraction of a physical law $y=y_o(x)$ from joint experimental data about $x$ and $y$ is treated. The joint, the marginal and the conditional probability density functions (PDF) are expressed by given data over an estimator whose kernel is the instrument scattering function. As an optimal estimator of $y_o(x)$ the conditional average is proposed. The analysis of its properties is based upon a new definition of prediction quality. The joint experimental information and the redundancy of  joint measurements are expressed by the relative entropy. With the number of experiments the redundancy on average increases, while the experimental information converges to a certain limit value. The difference between this limit value and the experimental information at a finite number of data represents the discrepancy between the experimentally determined and the true properties of the phenomenon. The sum of the discrepancy measure and the redundancy is utilized as a cost function. By its minimum a reasonable number  of data for the extraction of the law $y_o(x)$ is specified. The mutual information is defined by the marginal and the conditional PDFs of the variables. The ratio between mutual information and marginal information is used to indicate which variable is the independent one. The properties of the introduced statistics are demonstrated on deterministically and randomly related variables.
} 
\maketitle
\section{Introduction}
\label{intro}
The progress of natural sciences depends on advancement in the fields of experimental techniques and modeling of relations between experimental data in terms of physical laws.\cite{gs,les} By  utilizing computers a revolution appeared in the acquisition of experimental data while modeling still awaits a corresponding progress. For this purpose the modeling process should be generally described  in terms of operations that could be autonomously performed by a computer. A step in this direction was taken recently by a nonparametric statistical modeling of the probability distribution of measured data.\cite{ig} The nonparametric modeling requires no a priori assumptions about the probability density function (PDF) of measured data and therefore provides for a fairly general and autonomous experimental modeling of physical laws by a computer.\cite{gs,dh} Moreover, the inaccuracy of measurement caused by stochastic influences can be properly accounted for in the nonparametric modeling that further leads to the expression of experimental information, redundancy of repeated measurements and model cost function in terms of entropy of information. These variables have already been applied when formulating an optimal nonparametric modeling of PDF, in the most simple case of a one--dimensional variable.\cite{ig} However, more frequently than modeling of a PDF the problem is to extract a physical law from joint data about various variables and to analyze its properties. Therefore, the aim of this article is to propose a general statistical approach also to the solution of this problem. 

As an optimal statistical estimator of an experimental physical law we propose the conditional average (CA) that is determined by the conditional PDF.\cite{gs} This estimator represents a nonparametric regression whose structure is case independent; hence it can be generally programmed and autonomously determined by a computer.  Due to these convenient properties, we consider CA as a basis for the autonomous extraction of experimental physical laws  in data acquisition systems.

The fundamental steps of the proposed approach to extraction of experimental physical laws from given data are explained in the second section. We first define the estimators of the joint, the marginal and the conditional PDFs and derive from them the conditional average as an optimal estimator of a physical law that is hidden in joint data. In order to estimate the  number of data appropriate for the extraction of a physical law, we further introduce the statistics that characterize the information provided by joint measurements. In the third section of the article the properties of the CA estimator and the other introduced statistics are demonstrated on cases of deterministically and randomly related data.
 
\section{Statistics of joint measurements}
\label{section1} 

\subsection{Uncertainty of experimental observation}
Without loss of generality we consider a phenomenon  that can be quantitatively characterized by two scalar valued variables $x$ and $y$ comprising a vector $\vec{z}=(x,y)$.  We further assume that the phenomenon can be experimentally explored by repetition of joint measurements on a two--channel instrument having equal spans $S_x=(-L,L)$, $S_y=(-L,L)$. Their Cartesian product $S_{xy}=S_x \otimes S_y$ determines the joint span. We treat a measurement of a joint datum as a process in which the measured object generates the instrument output $\vec{z}=(x,y)$. The basic properties of the instrument and measurement procedure  can be characterized by a calibration based on a set of objects $\{\vec{w}_{kl}=(u_k, v_l); k=1,\ldots \; l=1,\ldots \}$ that represent joint physical units. Using these units,  a scale net can be determined in the joint span $S_{xy}$ of the instrument. In order to simplify the notation, we further omit the indices of units.

A common property of measurements is that the output of the instrument  fluctuates even when calibration is repeated.\cite{gs,les} We  describe this property by the joint PDF $\psi(\vec{z}|\vec{w})$, which characterizes the scattering of the instrument output at a given joint unit $\vec{w}$. For the sake of simplicity, we consider an instrument whose channels can be calibrated mutually independently. In this case the instrument scattering function is expressed by the product of scattering functions corresponding to both channels $\psi(\vec{z}|\vec{w})=\psi(x|u)\psi(y|v)$. Their mean values $u$, $v$, and standard deviations $\sigma_x$, $\sigma_y$ represent an element of the instrument scale and the scattering of instrument output at the joint calibration. These values can be estimated statistically by the sample mean and variance of both components measured during repeated calibration by a joint unit $\vec{w}$. The standard deviation $\sigma$ characterizes the uncertainty of the measurement procedure performed on a unit.\cite{gs,les} 
We further consider the most frequent case in which the output scattering does not depend on the channel index and the position $\vec{w}=(u,v)$ on the joint scale. In this case it can be expressed as a function of the difference $\vec{z}-\vec{w}=(x-u,y-v) $ and a common standard deviation $\sigma=\sigma_x=\sigma_y$ as $\psi(\vec{z}|\vec{w})=\psi(\vec{z}-\vec{w},\sigma)$. We consider scattering of instrument output during calibration as a consequence of random disturbances in the measurement system. When these disturbances are caused by contributions from mutually independent sources, the central limit theorem of the probability theory leads us to the Gaussian scattering function $\psi(\vec{z}-\vec{w},\sigma)={\rm g}(x-u,\sigma){\rm g}(y-v,\sigma)$, in which the scattering of a single component is determined by:   
\begin{equation}
\psi(x|u)\,={\rm g}(x-u,\sigma)\,=
\frac{1}{\sqrt{2\pi}\,\sigma}\exp \biggl[-\frac{(x-u)^2}{2\sigma}\biggr].
\end{equation}

\subsection{Estimation of probability density functions}
Let us consider a single measurement which yields a joint datum $\vec{z}_1=(x_1,y_1)$. We assume that this joint datum appears at the outputs of instrument channels, since it is the most probable at a given state $\vec{z}$ of the observed phenomenon and the instrument during measurement. Therefore, we utilize the measured datum $\vec{z}_1$ as the center of the probability distribution $\psi(\vec{z}-\vec{z}_1,\sigma)=\psi(x-x_1,\sigma)\psi(y-y_1,\sigma)$ that represents the corresponding state.

Consider next a series of $N$ repeated measurements which yield the basic data set $\{\vec{z}_i ;\,i=1,\ldots,N\}$. In accordance with the above--given interpretation of measured data we adapt to them the distributions $\{\psi(\vec{z}-\vec{z}_i,\sigma);\,i=1,\ldots,N\}$. If the data $\vec{z}_1,\ldots ,\vec{z}_N$ are spaced more than $\sigma$ apart, we assume that their scattering is caused by variation of the state $\vec{z}$ in repeated measurements and generally consider $\vec{z}$ as a random vector variable. Its joint PDF is determined by the statistical average over distributions $\{\psi(\vec{z}-\vec{z}_i,\sigma); i=1,\ldots,N\}$ as:
\begin{equation}\label{mm}
f_N(\vec{z})\,=\,\frac{1}{N}\,\sum_{i=1}^N \psi (\vec{z}-\vec{z}_i,\sigma) .
\label{pdfxy}
\end{equation}
This function represents an experimental model of PDF and resembles Parzen's kernel estimator, which is often used in statistical modeling of PDFs.\cite{par,dh} However, in Parzen's modeling the kernel width $\sigma$ plays the role of a smoothing parameter whose value decreases with the number of data $N$, which is not consistent with the general properties of measurements. In opposition to this, we consider $\sigma$ as an instrumental parameter that is determined by the inaccuracy of measurement.\cite{ig,dh}  In the majority of experimental observations $\sigma$ is a constant during measurements, and hence  need not be further indicated in the scattering function $\psi$.

From the joint PDF $f(\vec{z})=f(x,y)$ the marginal PDF $f(x)$ of a component $x$ is obtained by integration over the other component, for example:
\begin{equation}\label{pdfx}
f(x)\,=\,\int_{S_y} f(x,y) dy
\end{equation}
The conditional PDF of the variable $y$ at a given condition $x$ is then defined by the ratio of the joint PDF and the marginal PDF of the condition: 
\begin{equation}\label{cpdf}
f(y|x)\,=\,  \frac{f(x,y)}{f(x)}
\end{equation}
Using the experimental model of joint PDF (\ref{mm})  we obtain for the marginal and conditional PDFs the following kernel estimators:   
\begin{equation}\label{pdfxe}
f_N(x)\,=\,\frac{1}{N}\,\sum_{i=1}^N \psi (x-x_i,\sigma)
\end{equation}
\begin{equation}\label{cpdfe}
f_N(y|x)\,=\,\frac{\sum_{i=1}^N \psi (x-x_i,\sigma)\psi (y-y_i,\sigma)  }{\sum_{i=1}^N \psi (x-x_i,\sigma) }
\end{equation}

\subsection{Estimation of a physical law}
It is often observed that the joint PDF resembles a crest along some line $y=\hat{y}(x)$. We consider $\hat{y}(x)$ as an estimator of a hidden physical law $y=y_o(x)$ that provides for a prediction of a value $y$ from the given value $x$. If we repeat joint measurements, and consider only those that yield the value $x$, we can generally observe that corresponding values of the variable $y$ are scattered, at least due to the stochastic character of the measurements. As an optimal predictor of the variable $y$ at the given value $x$, we consider the value $\hat{y}$ that yields the minimum of the mean square prediction error $D$ at a given $x$:
\begin{equation}\label{D}
D\,=\,{\rm E} [(\hat{y} - y)^2|x]\, = \,\rm{min}(\hat{y})
\end{equation}
The minimum takes place when $\rm{d}D/\rm{d}\hat{y}=0$. The solution of this equation  yields as the optimal predictor $\hat{y}$ the conditional average 
\begin{equation}\label{CA}
\hat{y}(x)\,=\,{\rm E} [y|x]\,=\,\int_{S_y} y \,f(y|x) dy 
\end{equation}
By using Eq. \ref{cpdfe} for the conditional probability, we obtain for CA the superposition
\begin{equation}\label{CAN}
\hat{y}_N(x)\,=\,\frac{\sum_{i=1}^N y_i \psi (x-x_i,\sigma)}{\sum_{i=1}^N \psi (x-x_i,\sigma)}=\sum_{i=1}^N y_i C_i (x)
\end{equation}
The coefficients 
\begin{equation}\label{C}
C_i (x)\,=\,\frac{\psi (x-x_i,\sigma)}{\sum_{i=1}^N \psi (x-x_i,\sigma)}
\end{equation}
represent a normalized measure of similarity between the given value $x$ and sample values $x_i$ and satisfy the conditions:
\begin{equation}
\sum_{i=1}^N C_i (x)=1\,,
\end{equation}
\begin{equation}
0 \leq C_i (x) \leq 1.
\end{equation}
The more similar given value $x$ is to a datum $x_i$, the larger the coefficient $C_i(x)$ is and the contribution of the corresponding term $y_iC_i(x)$ to the sum in Eq.(\ref{CAN}). The prediction of the value $\hat{y}_N(x)$, which best corresponds to the given value $x$, thus resembles the associative recall of memorized items in the brains of intelligent beings, and therefore  could be treated as a basis for the development of computerized autonomous modelers of physical laws and related machine intelligence.\cite{gs}   

The predictor Eq. (\ref{CAN}) is completely determined by the set of measured data $\{\vec{z}-\vec{z}_i;\,i=1,\ldots,N\}$ and the instrument scattering function $\psi$. The predictor is not based on any {\em a priori} assumption about the functional relation between the variables $x$ and $y$, as is done for example when a physical law is described by  some regression function in which parameters are adapted to given data.  The conditional average Eq. (\ref{CAN}) can thus be treated as a nonparametric regression, although the scattering functions $\psi(\vec{z}-\vec{z}_i,\sigma)$ still depend on the parameters $\vec{z}_i,\sigma$. However, these parameters, as well as the form of the function $\psi$, are totally specified by measurements. They represent a property of the observed phenomenon and not an assumed auxiliary of the modeling. Since the form of the CA predictor does not depend on a specific phenomenon under consideration, it could be considered as a generally applicable basis for statistical modeling of physical laws in terms of experimental data in an autonomous computer. It is convenient  that Eq. (\ref{CAN}) can be simply generalized to a multi--dimensional case by substituting the condition and the estimated variable by the corresponding vectors.\cite{gs} Moreover, it is convenient that the ordering into dependent and independent variables is done automatically by a specification of the condition. 

\subsubsection{Description of predictor quality}
We can interpret a phenomenon which is characterized by the vector $\vec{z}=(x,y)$ as a process that maps the variable $x$ to the variable $y$. When the variables $x$ and $y$ are stochastic, we most generally describe this mapping by the joint PDF $f(x,y)$. Similarly, we can interpret the prediction of the variable ${\hat y}(x)$ from the given value $x$ as a process that runs in parallel with the observed phenomenon.  This process is also generally characterized by the PDF $f(x,{\hat y})$, while the relation between the variables $y$ and ${\hat y}$ is characterized by the PDF $f(y,{\hat y})$. The better the predictor is, the more the distribution  $f(y,{\hat y})$ is concentrated along the line $y= {\hat y}(x)$. For a good predictor we generally expect that the prediction error $E_r=y-{\hat y}$ is close to $0$. Since both variables are considered as stochastic ones, we expect that the first and second moments of the prediction error ${\rm E}[ y-{\hat y}]$, ${\rm E}[ (y-{\hat y})^2]$ are small, while for an exact prediction ${\rm E}[ y-{\hat y}]=0$, and ${\rm E}[ (y-{\hat y})^2]= 0$. 
The second moment of the error is equal to ${\rm E}[ (y-{\hat y})^2]={\rm Var} (y)+{\rm Var} ({\hat y})-2{\rm Cov} (y,{\hat y})+(m_y-m_{{\hat y}})^2$, where $m_y={\rm E}[y]$ and $m_{{\hat y}}={\rm E}[{\hat y}]$ denote mean values. If the variables $y$ and ${\hat y}$ are statistically independent and have equal mean values, the covariance vanishes: ${\rm Cov} (y,{\hat y})=0$, and $m_y-m_{{\hat y}}=0$, so that ${\rm E}[ (y-{\hat y})^2]={\rm Var} (y)+{\rm Var} ({\hat y})$. Based upon this property we introduce a relative statistic called the {\em predictor quality} with the formula
\begin{eqnarray}
Q&=&1-\frac{{\rm E}[ (y-{\hat y})^2]}{{\rm Var} (y)+{\rm Var} ({\hat y})} \nonumber \\
&=&\frac{2{\rm Cov} (y,{\hat y})}{{\rm Var} (y)+{\rm Var} ({\hat y})}-\frac{(m_y-m_{{\hat y}})^2}{{\rm Var} (y)+{\rm Var} ({\hat y})}
\end{eqnarray}
Its value equals $1$ for an exact prediction: ${\hat y}=y$, while it equals $0$, if the variables  $y$, ${\hat y}$ are statistically independent and have equal mean values. If the mean values differ: $m_y-m_{{\hat y}}\ne 0$, the quality $Q$ can also be negative. 

When the predictor is determined by the conditional average (\ref{CA}), we obtain for its mean value 
\begin{eqnarray}
m_{\hat y}={\rm E}[ {\hat y}]&=&\int {\hat y} f(x) dx =\int \int y f(y|x) f(x) dx dy  \nonumber \\
&=&\int \int y f(y,x) dx dy = {\rm E}[ y]=m_y.
\end{eqnarray}
Since in this case $m_y-m_{{\hat y}}=0$, we further get 
\begin{equation}
Q=\frac{2{\rm Cov} (y,{\hat y})}{{\rm Var} (y)+{\rm Var} ({\hat y})} 
\end {equation}
Similarly we get for the covariance 
\begin{eqnarray}
{\rm Cov} (y,{\hat y})&=&\int \int (y-m_y) ({\hat y}(x)-m_{ {\hat y}}(x)])f(y,x) dx dy \nonumber \\
&=&\int ({\hat y}(x)-m_{ {\hat y}}(x))(y-m_y) f(y|x) dy f(x) dx \nonumber \\
&=&\int ({\hat y}(x)-m_{{\hat y}}(x))^2 f(x) dx = {\rm Var} ({\hat y}),
\end{eqnarray}
so that the expected quality of the CA predictor is
\begin{equation}
Q=\frac{2{\rm Var} ({\hat y})}{{\rm Var} (y)+{\rm Var} ({\hat y})}.
\label{QCA} 
\end {equation}
In the case  when the relation between both components of the vector $\vec{z}$ is determined by some physical law $y_o(x)$, and only the measurement procedure introduces an additive noise $\nu$ with zero mean ${\rm E}[\nu]=0$, and variance ${\rm E}[\nu^2]=\sigma^2$, we can express  the variable $y$ as $y=y_o(x) +\nu$. In this case the following equations: ${\rm E}[(y-{\hat y})^2]=\sigma^2$, ${\rm Var} (y)={\rm Var} ({\hat y})+\sigma^2$  hold, and we get for the expected predictor quality the expression:
\begin{equation}
Q=\frac{2{\rm Var} ({\hat y})}{2{\rm Var} ({\hat y})+\sigma^2}.
\end {equation}
For ${\rm Var} ({\hat y})\gg\sigma^2/2$ we have $Q\approx 1$, while for ${\rm Var} ({\hat y})\ll\sigma^2/2$ we have $Q\approx 0$. In the last case ${\hat y}\approx {\rm constant}$, while $y$ fluctuates around this constant, and consequently  the prediction quality is low.
  
Since generally ${\rm Var} (y)\ge {\rm Var} ({\hat y})$ and ${\rm Var} ({\hat y})\ge 0$, we obtain from Eq. (\ref{QCA}) the inequality $0\le Q\le 1$. It describes a mean property, which need not be fulfilled exactly if the conditional average is statistically estimated from a finite number of samples $N$; but we can expect that it holds ever more with an increasing $N$. However, we can generally expect that with an increasing $N$, the statistically estimated CA ever better represents the underlying physical law $y=y_o(x)$. However, with an increasing $N$, the cost of experiments increases, and consequently  there generally appears the question: "How to specify a number of samples $N$ that is reasonable for the experimental estimation of a hidden law $y_o(x)$?" 

\subsection{Experimental information}
In order to answer the last question, we proceed with the description of the indeterminacy of the vector variable $\vec{z}$ in terms of the entropy of information. Following the definitions given for a scalar random variable in the previous article,\cite{ig} we first describe the indeterminacy of the component $x$. For this purpose we introduce a uniform reference PDF $\rho(x) = 1/(2L)$ that hypothetically corresponds to the most indeterminate noninformative observation of variable $x$; or to equivalently prepared initial states of the instrument before executing the experiments in a series of observations. By using this reference and the marginal PDF $f(x)$, we first define the indeterminacy of a continuous random variable by the negative value of the relative entropy\cite{ct,kol}
\begin{equation}
H_x=-\int_{S_x} f(x) \log \Bigl(\frac{f(x)}{\rho(x)}\Bigr) \,dx .
\label{Hz}
\end{equation}
Using the expressions for the reference, instrumental scattering function, and experimentally estimated PDF, we obtain the  expressions for the uncertainty $H_u$ of calibration performed on a unit $u$, the uncertainty $H_x$ of the component $x$, experimental information $I_x$ provided by $N$ measurements of $x$, and the redundancy $R_x$ of these measurements as follows \cite{ig}:
\begin{eqnarray}
H_u&=&-\int_{S_x} \psi(x,u)\log (\psi(x,u))\,dx - \log(2L), \nonumber\\
H_x&=&-\int_{S_x} f_N(x)\log (f_N(x))\,dx- \log(2L), \nonumber\\
I_x(N)&=&H_x-H_u,\nonumber\\
R_x(N)&=&\log (N) -I_x(N),
\label{infx}
\end{eqnarray}
Similar equations are obtained for the component $y$ by substituting $x\rightarrow y$. 

In order to describe the uncertainty of the random vector $\vec{z}$, we utilize the reference PDF that is uniform inside the joint span $S_{xy}$: $\rho(\vec{z}) = \rho(x)\rho(y)=1/(2L)^2$, and vanishes elsewhere. By analogy with the scalar variable we define the indeterminacy of the random vector $\vec{z}$ by the negative value of the relative entropy:\cite{ct}
\begin{equation}
H_{xy}=-\int\int_{S_{xy}} f(\vec{z}) \log \Bigl(\frac{f(\vec{z})}{\rho(\vec{z})}\Bigr) \,dxdy.
\label{Hz}
\end{equation}
In the case of a uniform reference PDF we obtain 
\begin{equation}
H_{xy}=-\int\int_{S_{xy}} f(\vec{z}) \log (f(\vec{z})) \,dx dy - 2\log (2L) .
\end{equation}
With this formula we then express the uncertainty of the joint instrument
calibration as 
\begin{equation}
H_\vec{w}=-\int\int_{S_{xy}} \psi(\vec{z},\vec{w})\log (\psi(\vec{z},\vec{w}))\,dx dy- 2\log (2L) .
\end{equation}
For $\sigma\ll L$ we obtain from the Gaussian scattering function $\psi(\vec{z},\vec{z}_i)={\rm g}(x-x_i,\sigma){\rm g}(y-y_i,\sigma)$ the approximation
\begin{equation}
H_\vec{w}\approx \log \Bigl(\frac{\sigma^2}{L^2}\Bigr)+\log \frac{\pi}{2}+1 ,
\end{equation}
The uncertainty of calibration depends on the ratio
between the scattering width $2\sigma$ and the instrument span $2L$ in both directions. The number
$2\log (\sigma / L)$ determines the lowest possible uncertainty of measurement
on the given two--channel instrument, as achieved at its joint calibration.

The indeterminacy of the random vector $\vec{z}$, which characterizes the scattering of
experimental data, is defined by the estimated joint PDF as 
\begin{equation}
H_{xy}=-\int\int_{S_{xy}} f_N(\vec{z})\log (f_N(\vec{z}))\,dx dy- 2\log (2L)
\end{equation}
and is generally greater than the uncertainty of calibration described by $H_\vec{w}$. Since $H_\vec{w}$ denotes the lowest possible indeterminacy of observation carried out over a given instrument, we define the joint 
experimental information $I_{xy}$ about vector $\vec{z}=(x,z)$ by the difference  
\begin{eqnarray}
I_{xy}(N)&=&H_{xy}-H_\vec{w}\nonumber \\
&=&-\int\int f_N(\vec{z}) \log (f_N(\vec{z}))\,dx dy \nonumber \\
&\phantom{=}&+\int\int \psi(\vec{z},\vec{w}) \log (\psi(\vec{z},\vec{w})) \,dx dy .
\label{infz}
\end{eqnarray}
Most properties of the uncertainty and information appertaining to a random vector are similar to those in the case of a scalar variable. For example, the reference density $\rho(\vec{z})$ can be arbitrarily selected since it is excluded from the specification of the experimental information.\cite{ig} Furthermore, the joint experimental information $I_{xy}(1)$ provided by a single measurement is zero. For a measurement which yields multiple samples
$\vec{z}_1, \ldots , \vec{z}_N$ that are mutually separated by several $\sigma$ in both directions,
the distributions $\psi(\vec{z},\vec{z}_1)={\rm g}(x-x_i,\sigma){\rm g}(y-y_i,\sigma)$ are nonoverlapping and the
first integral on the right of Eq.\,\ref{infz} can be approximated as 
\begin{eqnarray}
&-&\frac{1}{N}\sum_{i=1}^N\int\int\psi(\vec{z},\vec{z}_i) \log \Bigl[%
\frac{1}{N}\sum_{i=1}^N\psi(\vec{z},\vec{z}_i)\Bigr] \,dxdy  \nonumber \\
&\approx& \log (N)-\int\int \psi(\vec{z},\vec{z}_1) \log \psi(\vec{z},\vec{z}_1)\,dxdy
\end{eqnarray}
so that we get $I_{xy}(N)\approx\log (N)$. If the distributions $\psi(\vec{z},\vec{z}_i)$ are overlapping  but not concentrated at a single point, the inequality $0\leq I_{xy}(N)\leq \log (N)$ holds generally. Similarly as the entropy of information for a discrete random variable, the experimental information describes how much information is provided by $N$ experiments performed by an instrument that is not infinitely accurate.\cite{ct} In accordance with these properties the experimental information describes the complexity of experimental data in units of information entropy, which are here {\em nats}. 

When the distributions $\psi(\vec{z},\vec{z}_i)$ are nonoverlapping, $N$ repeated experiments yield the maximal possible information $\log (N)$. However, with an increasing number $N$, ever more overlapping of distributions $\psi(\vec{z},\vec{z}_i)$ takes place, and therefore  the experimental information $I_{xy}(N)$ increases more slowly than $\log ( N)$. Consequently, the repetition of joint measurements becomes on average ever more redundant with an increasing number $N$. The difference  
\begin{equation}
R_{xy}(N)=\log (N) -I_{xy}(N)\, .
\label{Rz}
\end{equation}
thus represents the redundancy of repeated joint measurements in $N$ experiments. Since the overlapping of distributions $\psi(\vec{z},\vec{z}_i)$ increases with an increasing number of experiments, the experimental information on average tends to a constant value $I_{xy}(\infty)$, and along with this, the redundancy increases with $N$. 

The number
\begin{equation}
K_{xy}(N)={\rm e}^{I_{xy}(N)}
\end{equation}
describes how many nonoverlapping distributions are needed to represent the experimental observation.
With an increasing $N$, the number $K_{xy}(N)$ tends to a fixed value $K_{xy}(\infty)$ that can be well estimated already from a finite number of experiments. We could conjecture  that $K_{xy}(\infty)$ approximately determines a reasonable number of experiments that provide sufficient data for an acceptable modeling of the joint PDF. However, it is still better to determine such a number from a properly introduced cost function of the experimental observation. With this aim we consider the difference $D_{xy}(N)=I_{xy}(\infty)-I_{xy}(N)$ as the measure of the discrepancy between the experimentally observed and the true properties of the phenomenon. An information cost function is then comprised of the redundancy and the discrepancy measure:
\begin{equation}
C_{xy}(N)=R_{xy}(N)+D_{xy}(N).
\end{equation}
Since the redundancy on average increases, while the discrepancy measure decreases with the number of measurements $N$, we expect that the cost function $C_{xy}(N)$ exhibits a minimum at a certain number $N_o$, which could be considered as an optimal one for the experimental modeling of a phenomenon. From the definition of redundancy and the discrepancy measure we further obtain $C_{xy}(N)=R_{xy}(N)+D_{xy}(N)=\log (N) - 2I_{xy}(N)+I_{xy}(\infty)$. Since the last term is a constant for a given phenomenon, it is not essential for the determination of $N_o$, and can be omitted from the definition of the cost function. This yields a more simple version
\begin{equation}
C_{xy}(N)=\log (N) - 2I_{xy}(N),
\end{equation}
which is more convenient for application since it does not include the limit value $I_{xy}(\infty)$. In a previous article \cite{ig} we have proposed a cost function that is comprised from the redundancy and the information measure of the discrepancy between the hypothetical and experimentally observed PDFs. However, such a definition is less convenient than the present one, although the values of $N_o$ determined from both cost functions do not differ essentially. Numerical investigations also show that the optimal number $N_o$ approximately corresponds to $K_{xy}(\infty)={\rm e}^{I_{xy}(\infty)}$ if the distribution of the data points is approximately uniform. 

Although the experimental information of a vector variable and its scalar components exhibits similar properties, their values generally do not coincide since the overlapping of distributions  $\psi(\vec{z},\vec{z}_i)$ generally differs from that of distributions $\psi(x,x_i)$ or $\psi(y,y_i)$. Therefore, the experimental information provided by joint measurements generally differs from that provided by measurements of single components.

\subsection{Mutual information and determination of one variable by the other}

In order to describe the information corresponding to the relation between variables $x,y$ we introduce conditional entropy. At a given value $x$ we express the entropy pertaining to the variable $y$ by the conditional PDF as 
\begin{equation}
H_{y|x}=-\int_{S_y} f(y|x)\log \Bigl(\frac{f(y|x)}{\rho(y)}\Bigr)\,dy
\label{infc}
\end{equation}
If we express in Eq. (\ref{Hz}) the joint PDF by the conditional one $f(\vec{z})=f(y|x)f(x)$ we obtain the following equation:
\begin{equation}
H_{xy}=\overline{H_{y|x}} + H_x
\end{equation}
in which $\overline{H_{y|x}}$ denotes the average conditional entropy of information
\begin{equation}
\overline{H_{y|x}}=-\int_{S_x} H_{y|x} f(x) \,dx .
\end{equation}
When we exchange the meaning of the variables we get  
\begin{equation}
H_{xy}=\overline{H_{x|y}} + H_y.
\end{equation}
Based on these equations and Eq. (\ref{infz}) we obtain the following relation between the joint and the conditional information 
\begin{eqnarray}
I_{xy}&=&\overline{H_{x|y}} + H_y-H_u-H_v \nonumber\\
&=&\overline{I_{y|x}} + I_x =\overline{I_{x|y}} + I_y
\end{eqnarray}
where the conditional information is defined by
\begin{eqnarray}
\overline{I_{x|y}}=\overline{H_{x|y}} - H_u \quad {\rm or} \quad \overline{I_{y|x}}=\overline{H_{y|x}} - H_v.
\end{eqnarray}

When the components of the vector $\vec{z}$ are statistically independent, the joint PDF is equal to the product of marginal probabilities and the joint information is given by the sum $I_{xy}=I_x +I_y$, which represents the maximal possible information that could be provided by joint measurements. However, when $x$ and $y$ are not statistically independent,  the joint information is less than the maximal possible one: $I_{xy}<I_x +I_y$. The difference 
\begin{equation}
I_m=I_x +I_y - I_{xy}=I_x-\overline{I_{x|y}}= I_y-\overline{I_{y|x}}.
\label{Im}
\end{equation}
can be interpreted as the experimental information that a measurement of one variable provides about another one and is consequently called the mutual information.\cite{ct,cla,haop,hau} In accordance with the previous interpretation of the redundancy, it follows from the last two terms in Eq. (\ref{Im}) that  the mutual information also describes how redundant on average is a measurement of the variable $y$ at a given $x$ or vice versa. In accordance with the definition of the redundancy of a certain number $N$ of measurements $R_x(N)=\log (N) - I_x$, we further define also the mutual redundancy of $N$ joint measurements 
\begin{equation}
R_m(N)=\log (N) - I_m (N)\,.
\label{Rm}
\end{equation}
If we then take into account all the definitions of the redundancies and types of information, we obtain the formula:
\begin{equation}
R_{xy}(N)=R_x(N) +R_y(N) - R_m(N)
\label{Rall}
\end{equation}
It should be pointed out that redundancies $R_{xy}(N)$, $R_x(N)$, $R_y(N)$, and $R_m(N)$ generally increase with $N$, while the corresponding experimental information tends to fixed values that correspond to the amount of data needed for presenting related variables.

In order to describe quantitatively how well determined the value of the variable $y$ by the value of $x$ is on average, we propose a {\em relative measure of determination} by the ratio
\begin{equation}
\overline{D_{y|x}}=\frac{I_m}{I_y} = 1-\frac{\overline{I_{y|x}}}{I_y} .
\label{Det}
\end{equation}
If $\overline{D_{y|x}}>\overline{D_{x|y}}$, the value of the variable $x$ better determines the value of $y$ than vice versa. In this case the variable $x$ could be considered as more fundamental for the description of the phenomenon, and consequently as an independent one. In the case of functional dependence described by a physical law $y=y_o(x)$, the relative measure of determination is $\overline{D_{y|x}}=1$, while for the statistically independent variables $x$ and $y$ it is $\overline{D_{y|x}}=0$.

The entropy of information is generally decreased if the distribution of scattered experimental data at a given $x$ is compressed to the estimated physical law $\hat{y}(x)$. The corresponding information gain is in drastic contrast to the information loss that is caused by the noise in a measurement system.\cite{sha}

\section{Illustration of statistics}
\label{section2}
 
\subsection{Data with a hidden law}

\begin{figure}
\centering
\includegraphics[width=3.375in]{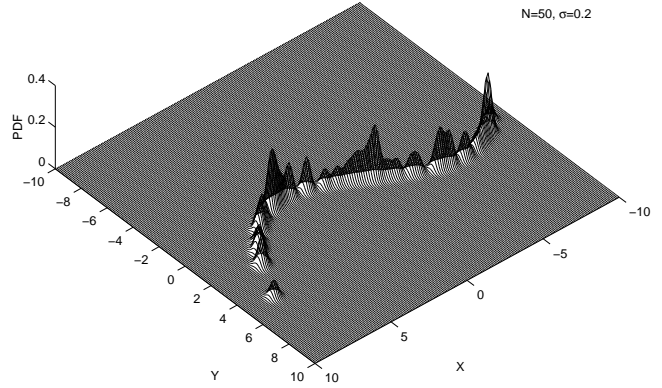}
\caption{The joint PDF $f(x,y)$ utilized to demonstrate the properties of the conditional average predictor.}
\label{figpdfz}
\end{figure}

\begin{figure}
\centering 
\includegraphics[width=3.375in]{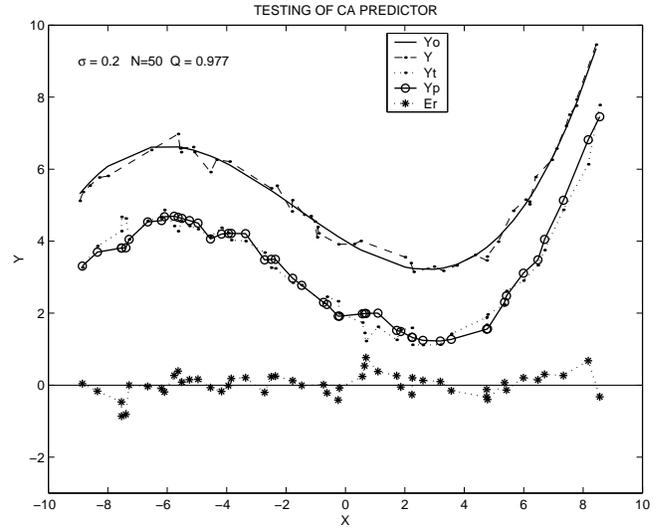}
\caption{Testing of CA predictor. Curves representing the underlying law and given data $y_o,y$ -- (top), test and predicted  data $y_t,y_p$ -- (middle), and prediction error $E_r=y_p-y_t$ -- (bottom) are displaced in vertical direction for a better visualization.}
\label{figCAN}
\end{figure}

The purpose of this section is to demonstrate graphically the basic properties of the statistics introduced above. For this purpose it is most convenient  to generate data numerically since in this case the relation between the variables $x$ and $y$, as well as the properties of the scattering function $\psi (\vec{z})$, can be simply set. For our demonstration we arbitrarily selected a third order polynomial law $y_o(x)=[x(x-5)(x+10)]/100$ and the Gaussian scattering function with standard deviation $\sigma=0.2$. To simulate the basic data set $\{x_i, y_i ;\,i=1,\ldots ,N\}$,  we first calculated $50$ sample values $x_i$ by summing two random terms obtained from a generator with a uniform distribution in the interval $[-8,+8]$ and from a Gaussian generator having the mean value $0$ and standard deviation $\sigma=0.2$. The corresponding sample values $y_i$ were then calculated as a sum of terms obtained from the selected law $y_o(x_i)$ and the same random Gaussian generator with a different seed. The generated data $\{x_i, y_i ;\,i=1,\ldots,50\}$ were used as centers of scattering function when estimating the joint PDF based on Eq.\,(\ref {pdfxy}). An example of such PDF is shown in Fig.\,\ref{figpdfz}, while the corresponding  joint data of the basic set are shown by points in the top curve of Fig.\,\ref{figCAN} together with the underlying law $y_o(x)$. 

The conditional average predictor, which corresponds to the presented example, was modeled by inserting data from the basic data set into Eq.\,(\ref{CAN}). To demonstrate its performance, we additionally generated a test data set by the same procedure as in the case of the basic data set, but with different seeds of all the random generators. Using the values $x_{i,t}$ of the test set, we then predicted the corresponding values $\hat{y}_i$ by the modeled CA predictor. With this procedure we simulated a situation that is normally met when a natural law is modeled and tested based upon experimental data. The test and predicted data are shown by the middle two curves in Fig.\,\ref{figCAN}. From both data sets the prediction error $E_r=\hat{y}-y_{t}$ was calculated that is presented by the bottom curve (..*..) in Fig.\,\ref{figCAN}. The curve representing the predicted data (--o--) is smoother than the curve representing the original test data (..$\cdot$..). This property is a consequence of smoothing caused by estimating  the conditional mean value from various data included in the modeled CA predictor. In spite of this smoothing, it is obvious that the characteristic properties of the relation between the variables $x$ and $y$ is approximately extracted from the given data by the CA predictor. This further means that the properties of the hidden law $y=y_o (x)$ can be approximately described in the region where measured data appear based on a finite number of joint samples. 

\begin{figure}
\centering 
\includegraphics[width=3.375in]{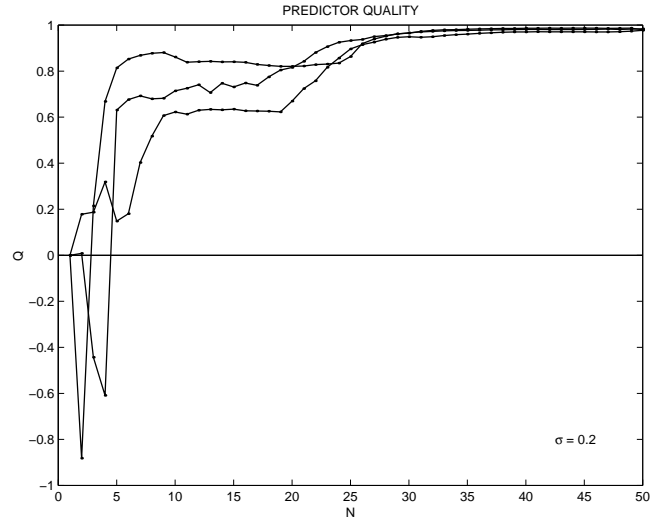}
\caption{Dependence of predictor quality $Q$ on number of samples $N$ determined by various statistical data sets.}
\label{figCAQ}
\end{figure}

The quality of estimation of the hidden law $y_o(x)$ depends on the values and number $N$ of statistical samples utilized in Eq.\,(\ref{CAN}) in the modeling of CA and its testing. To demonstrate this property, we repeated the complete procedure three times, using various statistical data sets with increasing $N$ and determined the dependence of predictor quality $Q$ on $N$. The result is presented in Fig.\,\ref{figCAQ}. 
The quality statistically fluctuates with the increasing $N$, but the fluctuations are ever less pronounced, so that quality determined from different data sets converges to a common limit value at a large $N$. In our example with $\sigma =0.2$ the limit value is approximately $Q=0.98$. With increasing $N$, the curves corresponding to different data sets join approximately at $N_{CA}\approx 30$. At a higher $N$ the fluctuations of $Q$ are ever less expressive. We could conjecture  that about $30$ data values are needed to model the CA predictor in the presented case approximately.  

The smaller the scattering width $\sigma$ is, the higher generally the limit value of the predictor quality is, but on average $Q$ is still less than $1$ if $1/\sigma$ and $N$ are finite. This property is in tune with the well--known fact that it is impossible to determine exactly the law $y=y_o(x)$ from joint data that are measured by an instrument which is subject to output scattering due to inherent stochastic disturbances. 

\begin{figure}
\centering 
\includegraphics[width=3.375in]{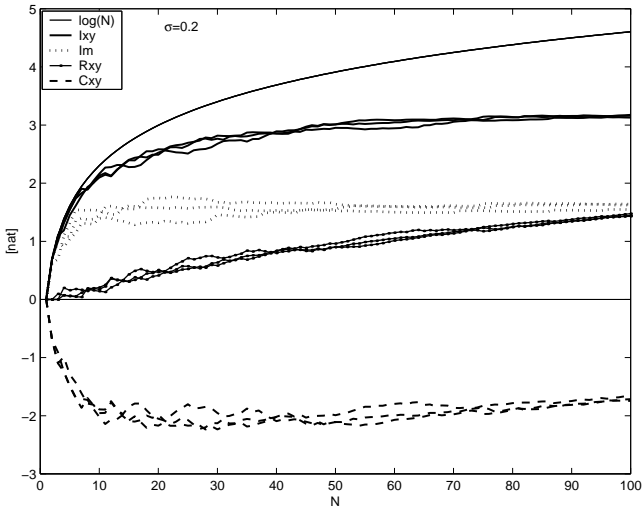}
\caption{Dependence of $\log (N)$, experimental information $I_{xy}$, mutual information $I_{m}$, redundancy $R_{xy}$, and cost function $C_{xy}$ on the number of samples $N$ determined by various statistical data sets.}
\label{figIxyz}
\end{figure}

The properties of the statistics  that are formulated based upon the entropy of information  are demonstrated for the case with $\sigma =0.2$ in Fig.\,\ref{figIxyz}. 
It shows the
dependence of experimental information $I_{xy}$, mutual information $I_{m}$, redundancy $R_{xy}$, and cost function $C_{xy}$ on the number of samples $N$ for three different sample sets. In the same figure the maximal possible information, which corresponds to the ideal case with no scattering, is also presented by the curve $\log (N)$, since it represents the basis for defining the redundancy. Similarly as in the one--dimensional case \cite{ig}, the experimental information $I_{xy}$ in the two--dimensional case also converges with increasing $N$ to a fixed value. In the presented case the limit value is $I_{xy}(\infty )\approx 3.2$, which yields the number $K_\infty\approx 25$. This number is approximately equal to the ratio of standard deviation of variable $x$ and the scattering width $\sigma$ and describes how many uniformly distributed samples are needed to represent the PDF of the data.\cite{ig} Due to the convergence of experimental information to a fixed value, the curve $I_{xy}(N)$ starts to deviate from $\log (N)$ with the increasing $N$. Consequently the redundancy $R_{xy}=\log (N)-I_{xy}(N)$ starts to increase, which further leads to the minimum of the cost function  $C_{xy}(N)=\log (N)-2I_{xy}(N)$. The minimum is not well pronounced due to statistical variations, but it takes place at approximately $N_o\approx 30$. Not surprisingly, the optimal number $N_o$ approximately corresponds to $K_\infty$ and also to $N_{CA}$. 

\begin{figure}
\centering 
\includegraphics[width=3.375in]{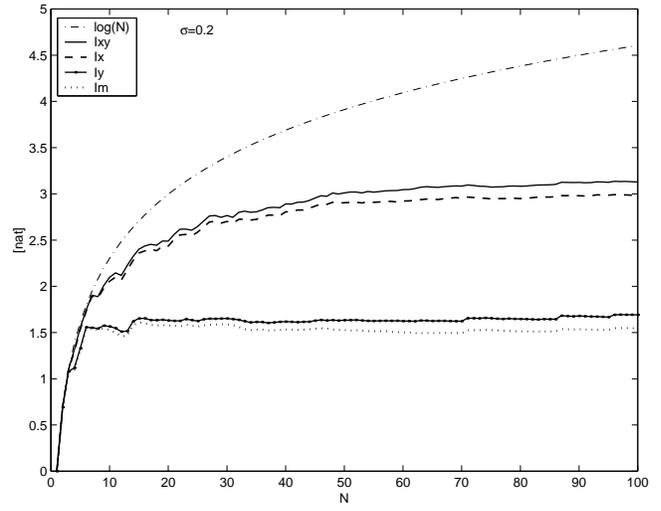}
\caption{Dependence of $\log (N)$, experimental information $I_{xy}$, marginal informations $I_{x},I_{y}$, and mutual information $I_{m}$ on the number of samples $N$.}
\label{figIxIyz}
\end{figure}

Similarly as the joint experimental information $I_{xy}$, the marginal experimental information $I_{x}$, $I_{y}$ also converges to fixed values with increasing $N$.\cite{ig} These statistics are presented in Fig.\,\ref{figIxIyz} for the same data generator as applied in the case of Fig.\,\ref{figIxyz}. 
The sample values of variable $x$ take place in a larger interval than those of  variable $y$. Hence there is less overlapping of scattering functions comprising the marginal PDF of $x$ and consequently $I_x$ is larger than $I_y$. It is also characteristic that $I_{xy}$ is larger than $I_x$ since the data points in the joint span $S_{xy}$ are more separated than in the marginal span $S_{x}$. Since the mutual information $I_m$ is defined as $I_m=I_x + I_y  - I_{xy}$, its properties depend on both the marginal and the joint information, and consequently $I_m $ converges more quickly to the limit value than the experimental information $I_{xy}$.

\begin{figure}
\centering 
\includegraphics[width=3.375in]{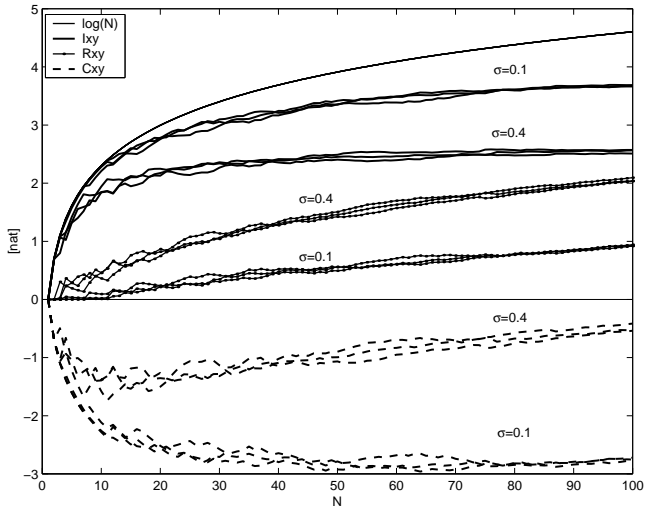}
\caption{Dependence of $\log (N)$, experimental information $I_{xy}$, redundancy $R_{xy}$, and cost function $C_{xy}$ on the number of samples $N$ determined from various data sets and scattering widths $\sigma$.}
\label{figIxyz0104}
\end{figure}

To demonstrate the influence of scattering width on the presented statistics the calculations were repeated with  $\sigma=0.1$ and $0.4$. The results are presented in Fig.\,\ref{figIxyz0104}. For the sake of clear presentation, the curves representing the mutual information $I_{m}$ are omitted. As could be expected, the limit value of $I_{xy}$ increases with decreasing $\sigma$. This property is consistent with the well--known fact that more information can be obtained by experimental observation when using an instrument of higher accuracy that corresponds to a lesser scattering width. In opposition to this, the redundancy of measurement decreases, and along with it, the optimal number $N_o$ increases with the decreasing scattering width.  

\begin{figure}
\centering 
\includegraphics[width=3.375in]{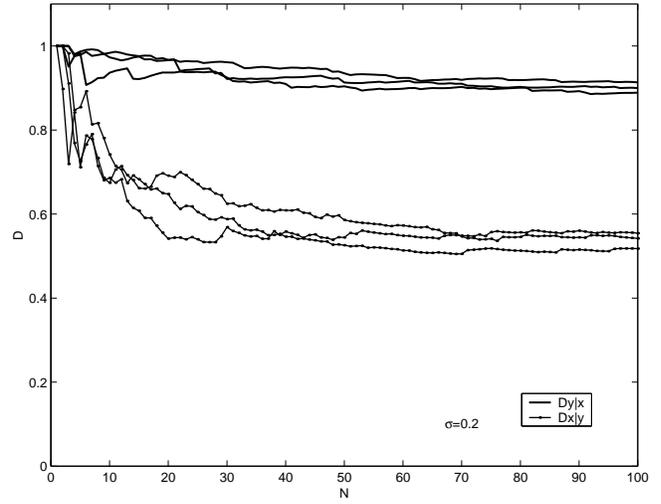}
\caption{Dependence of relative measure of determination $\overline{D_{y|x}}$ -- (top lines) and $\overline{D_{x|y}}$ -- (bottom lines) on the number of samples $N$ determined from various statistical data sets.}
\label{figDz}
\end{figure}

From the calculated mutual and marginal information, the relative measures of determination $\overline{D_{y|x}}$ and $\overline{D_{x|y}}$ were further determined using various statistical data sets. The results are presented in Fig.\,\ref{figDz} for the case of scattering width $\sigma=0.2$. 
When the number of data $N$ surpasses the interval around the optimal number $N_o$, statistical variations of $\overline{D_{y|x}}$ and $\overline{D_{x|y}}$ become less pronounced and their values settle close to limit ones. The limit value $\overline{D_{x|y}}$ is essentially lower  than $\overline{D_{y|x}}$. This is the consequence of the fact that in our case the variable $y$ is uniquely determined by the underlying law $y_o(x)$ based upon the variable $x$, but not vice versa. In our case, there are three values of the variable $x$ corresponding to a value of $y$ in a certain interval. Consequently, $y$ is better determined by a given $x$ than vice versa, which further yields $\overline{D_{y|x}}>\overline{D_{x|y}}$. Hence the relative measure of determination indicates that variable $x$ could be considered more fundamental for the description of the relation between the variables $x$ and $y$. 

\subsection{Data without a hidden law}

\begin{figure}
\centering
\includegraphics[width=3.375in]{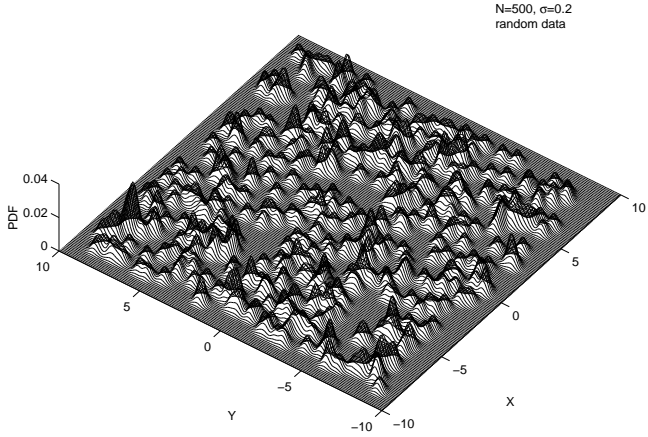}
\caption{The joint PDF $f(x,y)$  of $N=500$ statistically independent random data with $\sigma=0.2$.}
\label{figpdfr}
\end{figure}

\begin{figure}
\centering 
\includegraphics[width=3.375in]{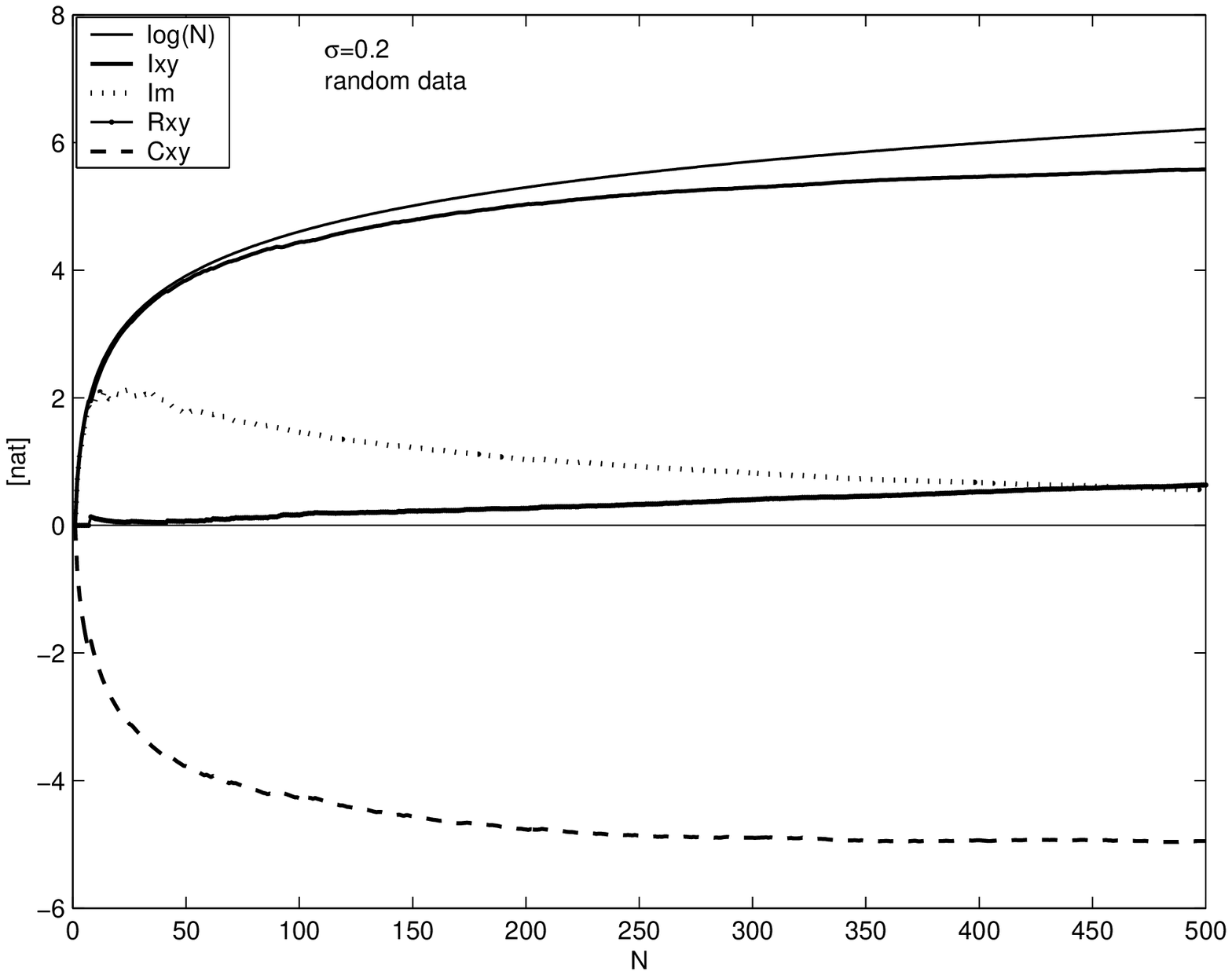}
\caption{Dependence of $\log (N)$, experimental information $I_{xy}$, redundancy $R_{xy}$, and cost function $C_{xy}$ on the number of samples $N$ determined by various statistical data sets and scattering widths $\sigma$.}
\label{figIxyr}
\end{figure}

\begin{figure}
\centering 
\includegraphics[width=3.375in]{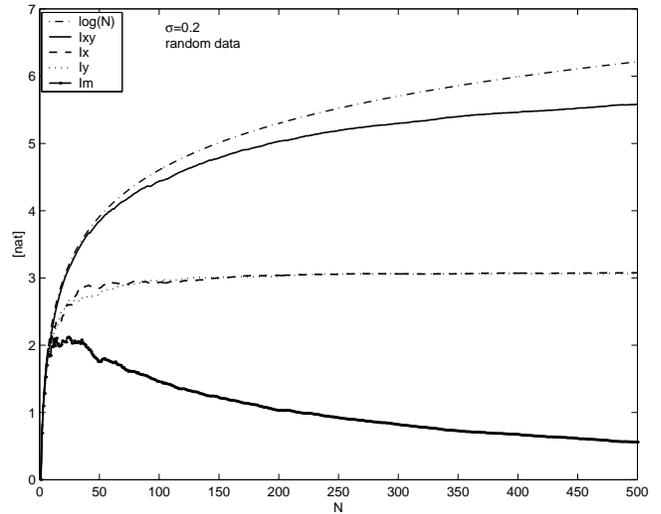}
\caption{Dependence of $\log (N)$, experimental information $I_{xy}$, marginal informations $I_{x},I_{y}$, and mutual information $I_{m}$ on the number of samples $N$ in the case of statistically independent random variables $x,y$.}
\label{figIxIyr}
\end{figure}

\begin{figure}
\centering 
\includegraphics[width=3.375in]{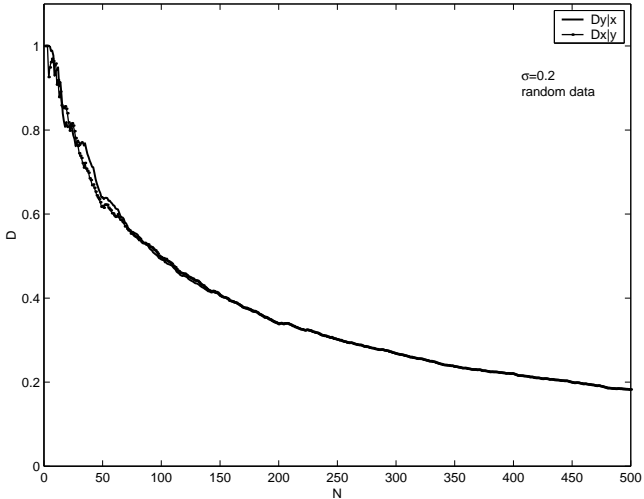}
\caption{Dependence of relative measure of determination $\overline{D_{y|x}}$ -- (top lines) and $\overline{D_{x|y}}$ -- (bottom lines) on the number of random samples $N$ in the case of statistically independent random data with  $\sigma=0.2$.}
\label{figDr}
\end{figure}

To support the last conclusion let us examine an example in which the sample values of the variables $x$ and $y$ were calculated by two statistically independent random generators. The corresponding joint PDF is shown in Fig.\,\ref{figpdfr}, while the properties of the other statistics are demonstrated by Figs.\,\ref{figIxyr}, \ref{figIxIyr} and \ref{figDr}. 

The properties of the presented statistics could be understood, if the overlapping of scattering functions comprising the estimator of the joint PDF is examined. In the previous case with the underlying law $y_o(x)$, the joint data are distributed along the corresponding line where $-8\le x\le +8$, while in the last case, they take place in the square region $-8\le x\le +8, -8\le y\le +8$. Consequently, the number of samples with nonoverlapping scattering functions  in the last case is approximately $L/\sigma=16$ times larger than in the previous case. In the last case we can therefore expect the optimal number of samples in the interval around $N_{ro}\approx 16N_o=480$. Since in the last case a larger region is covered by the joint PDF, the overlapping of scattering functions is less probable than previously, and therefore, the joint experimental information $I_{xy}$ deviates less quickly from the line $\log(N)$ with the increasing $N$. Therefore, the redundancy increases less quickly and the minimum of the cost function takes place at a much higher number of $N_{ro}= 480$, which corresponds well to our estimation. Since in the last case the experimental information $I_{xy}$ converges less quickly to the limit value than the marginal information $I_{x},I_{y}$, the mutual information $I_{m}$ first increases and later decreases to its limit value. Related to this is the approach of relative measures of determination $\overline{D_{y|x}},\overline{D_{x|y}}$ to much lower limit values as in the previous case. Since the marginal information $I_{x},I_{y}$ is approximately equal, the  curves representing $\overline{D_{y|x}},\overline{D_{x|y}}$ join with increasing $N$, and there is no argument to consider any variable as a more fundamental one for the description of the phenomenon under examination. This conclusion is consistent with the fact that the centers of the scattering functions are determined by two statistically independent random generators. However, the limit values of the statistics $\overline{D_{y|x}},\overline{D_{x|y}}$ are not equal to zero since the region $-8\le x\le +8, -8\le y\le +8$ where the data appear is limited, while the characteristic region $-\sigma\le x\le +\sigma, -\sigma\le y\le +\sigma$ covered by the joint scattering function does not vanish. 

\section{Conclusions}
\label{section3}

Following the procedures proposed in the previous article \cite{ig}, we have shown how the joint PDF of a vector variable $\vec{z}=(x,y)$ can be estimated nonparametrically based upon measured data. For this purpose the inaccuracy of joint measurements was considered by including the scattering function in the estimator. It is essential that the properties of the scattering function need not be a priori specified, but could be determined experimentally based upon calibration procedure. 
The joint PDF was then transformed into the conditional PDF that provides for an extraction of the law $y_o(x)$ that relates the measured variables $x,y$. For this purpose the estimation by the conditional average $y_o(x)\approx{\rm E} [y|x]$ is proposed. The quality of the prediction by the conditional average is described in terms of the estimation error and the variance of the measured data. It is outstanding that the quality exhibits a convergence to some limit value that represents the measure of applicability of the proposed approach. Examination of the quality convergence makes it feasible to estimate  an appropriate number of joint data needed for the modeling of the law. It is important that the conditional average makes feasible a nonparametric autonomous extraction of underlying law from the measured data. 

Using the joint PDF estimator we have also defined the experimental information, the redundancy of measurement and the cost function of experimental exploration. It is characteristic that experimental information converges with an increasing number of joint samples to a certain limit value which characterizes the number of nonoverlapping scattering distributions in the estimator of the joint PDF. The most essential terms of the cost function are the experimental information and the redundancy. During cost minimization the experimental information provides for a proper adaptation of the joint PDF model to the experimental data, while the redundancy prevents an excessive growth of the number of experiments. By the position of the cost function minimum we introduced the optimal number of the data that is needed to represent the phenomenon under exploration. This number roughly corresponds to the ratio between the magnitude of the characteristic region where joint data appear and the magnitude of the characteristic region covered by the joint scattering function. It also corresponds to the appropriate number estimated from the quality of prediction by the conditional average. Based upon the  experimental information corresponding to the joint and marginal PDFs, the mutual information has been introduced and further utilized in the definition of the relative measure of determination of one variable by another. This statistic provides an argument for considering one variable as a fundamental one for the description of the phenomenon. 

In this article we graphically present the properties of the proposed statistics by two characteristic examples that represent data related by a certain law and statistically independent random data. The exhibited properties agree well with the expectations given by experimental science. The problems related to the extraction of laws representing relations such as $y^2+x^2=1$ and the expression of physical laws by differential equations or analytical modeling were not considered. For this purpose the statistical methods are developed in the fields of pattern recognition, system identification and artificial intelligence. 

\noindent{\bf Acknowledgment}
\newline\noindent The research was supported by the Ministry of Science and Technology of Slovenia and EU COST.

\end{document}